\documentclass
[aps,prl,twocolumn,floatfix,english,showpacs,10pt,superscriptaddress]{revtex4-2}%
\usepackage{graphicx}
\usepackage{booktabs}
\usepackage{array}
\usepackage{makecell}
\usepackage{amsmath}
\usepackage{physics}
\usepackage{amssymb}
\usepackage{colordvi}
\usepackage{verbatim}
\usepackage{xcolor}
\usepackage{mathrsfs}
\usepackage{epsfig}
\usepackage{lipsum}
\usepackage{amsfonts}
\usepackage{makecell}
\usepackage{esint}
\usepackage{bm}
\usepackage[most]{tcolorbox}

\usepackage[unicode=true, breaklinks=false, pdfborder={0 0 1}, backref=false,
colorlinks=true, linkcolor=blue, urlcolor=blue, citecolor=blue]{hyperref}%
\providecommand{\U}[1]{\protect\rule{.1in}{.1in}}
\providecommand{\U}[1]{\protect\rule{.1in}{.1in}}
\setcitestyle{numbers,square}

\begin{document}

\title{Scattering-Induced Magnon Layer-Hall Transport beyond Band Geometry}

\author{Zhiping Xue}
\affiliation{School of Physics, Huazhong University of Science and Technology, Wuhan 430074, China}

\author{Zhoujian Sun}
\affiliation{State Key Laboratory of Quantum Functional Materials, Department of Physics, and Guangdong Basic Research Center of Excellence for Quantum Science, Southern University of Science and Technology (SUSTech), Shenzhen 518055, China}

\author{Xiyin Ye}
\affiliation{School of Physics, Huazhong University of Science and Technology, Wuhan 430074, China}

\author{Lei Zhang}
\affiliation{State Key Laboratory of Quantum Optics Technologies and Devices, Institute of Laser Spectroscopy,
Shanxi University, Taiyuan 030006, China}
\affiliation{Collaborative Innovation Center of Extreme Optics, Shanxi University, Taiyuan 030006, China}

\author{Tao Yu}
\email{taoyuphy@hust.edu.cn}
\affiliation{School of Physics, Huazhong University of Science and Technology, Wuhan 430074, China}

\date{\today}

\begin{abstract}

The layer Hall effect has been exclusively attributed to layer-locked Berry curvature, posing a fundamental barrier to its realization in conventional magnets. Here we report a fundamentally distinct layer Hall effect for bosonic excitations, i.e., magnons, which originates solely from non-reciprocal dipolar scattering at heterointerfaces, thereby decoupling the phenomenon from geometric-phase mechanisms. Using a microscopic scattering theory, we demonstrate that a longitudinal temperature gradient drives opposite transverse thermal Hall currents in a nanowire atop a magnetic film, with the direction fully reconfigurable by the applied magnetic field. The effect yields a significant Hall angle of $\sim 6^{\circ}$ in conventional magnetic heterostructures, eliminating the need for topological engineering. Our findings establish a scattering-driven paradigm for layer Hall effect, extendable to ferrons and polar phonons, and predict a Hall response that is readily detectable in conventional magnetic heterostructures.

\end{abstract}

\maketitle

The layer Hall effect (LHE) describes a Hall response where electrons in different layers are spontaneously deflected in opposite transverse directions by layer-locked Berry curvature. It was first experimentally observed in the even-layered antiferromagnetic topological insulator ${\rm MnBi_2Te_4}$~\cite{Gao2021} by applying a perpendicular electric field that breaks space-time inversion symmetry and activates nonzero layer-resolved anomalous Hall conductance. Beyond ${\rm MnBi_2Te_4}$, LHE has been theoretically proposed in various platforms, including related materials~\cite{Gao2021,Xu2024,Chen2024,Dai2022,Peng2023}, ferroelectrics~\cite{Zhang2023,Zhang2024}, multiferroic van der Waals bilayers~\cite{Feng2023,Liu2024}, and transition metal oxides~\cite{Tao2024}. Encoding Berry curvature in the layer degree of freedom, LHE shows potential for topological antiferromagnetic spintronics and next-generation low-power, electrically controllable devices.

Charge-neutral quasiparticles, including photons~\cite{Imbert1972,Bliokh2013,OConnor2014,Cai2017,Shah2021,Zhao2021,Gao2021Continuous}, phonons~\cite{Strohm2005,Inyushkin2007b,Mori2014,Sheng2006,Zhang2010,Qin2012,Sun2020,Z.Gao2021,Lefrancois2022,Shragai2026}, and magnons~\cite{Onose2010, Katsura2010, Matsumoto2011b, Matsumoto2011, Ideue2012, Shindou2013b, Shindou2013, Matsumoto2014, Mook2014, Cao2015, Owerre2016, Owerre2017, Gunnink2021, Choi2023, deOliveira2023,Chisnell2015,Mook2019,Zhang2019,Zhou2026}, exhibit the Hall effect owing to nontrivial band geometry. However, unlike the LHE where Berry curvature is locked to specific layers, the Berry curvature in these bosonic systems is generally delocalized across the lattice. In magnon systems~\cite{Serga2010,Lenk2011,Bauer2012,Chumak2015,Grundler2016,Demidov2017,Manchon2019,Yu2024}, this transverse deflection typically necessitates materials with strong spin-orbit coupling~\cite{Dzyaloshinsky1958,Moriya1960}. Enhancing the Berry curvature in ferromagnetic insulators like yttrium iron garnet (YIG) generally presents a challenge since it requires deliberate structural engineering, such as the fabrication of nanostructured magnetic moir\'e lattices~\cite{Wang2023,Zhang2013,Elyasi2019,Mook2014b,Owerre2016JP,Kim2016,Hirosawa2020,Kondo2021,Wang2020b,Bostrom2023,Shindou2013,Li2018}. While effective, these approaches often introduce additional damping that impedes wave propagation.

\begin{figure}[htp!]	
\centering
\includegraphics[width=0.5\textwidth,trim=0.0cm 0cm 0cm 0.0cm]{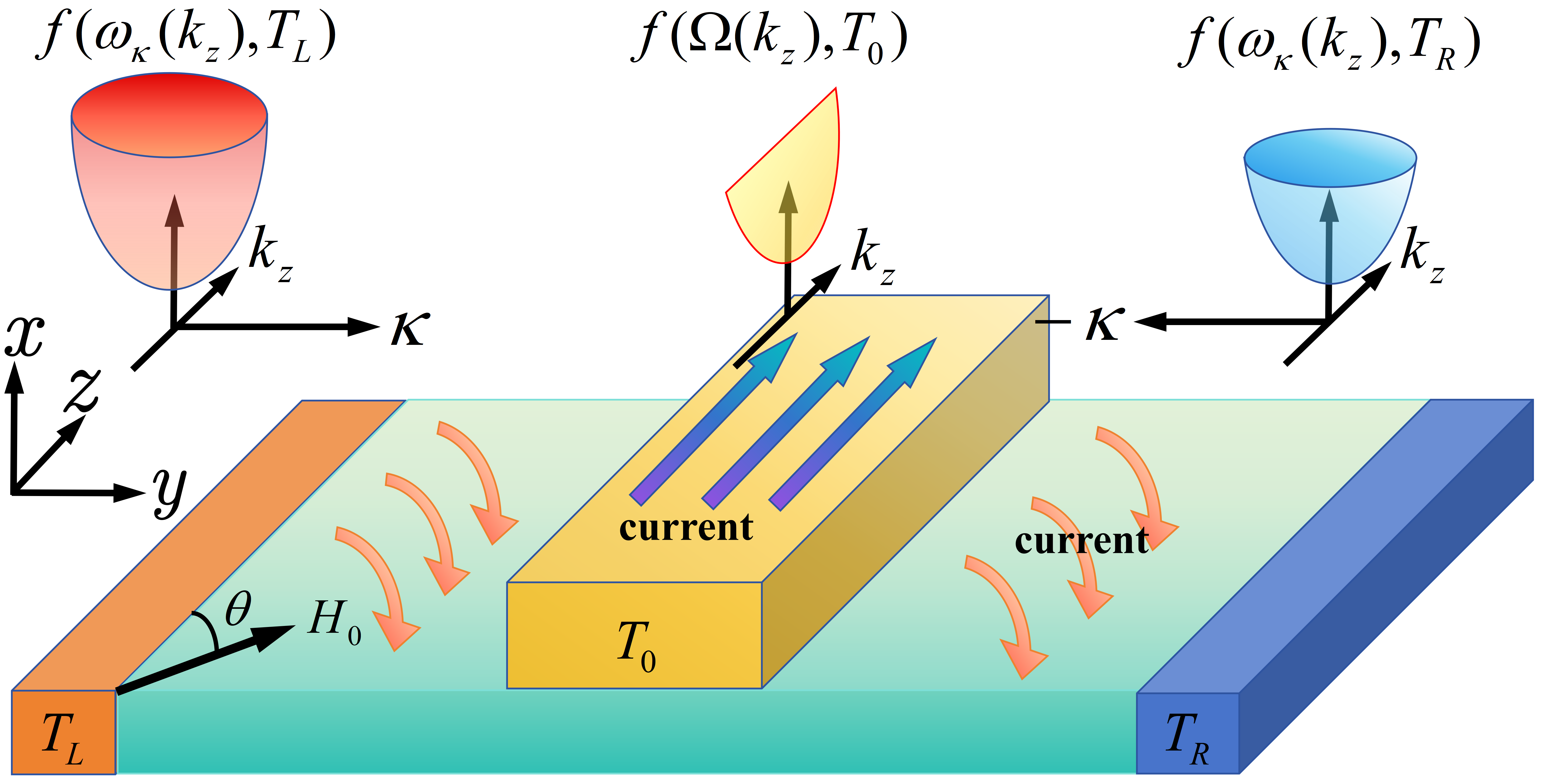}
\caption{Magnon layer-Hall transport driven by dipolar scattering. The hybrid system consists of a magnetic nanowire (width $w$, thickness $d$) placed atop a magnetic film (thickness $s$). The magnetization of the film (nanowire) is tilted by an angle $\theta$ $(\tilde{\theta})$ relative to the wire $\hat{\bf z}$-axis. The film is sandwiched between two thermal reservoirs at temperatures $T_L$ and $T_R$, establishing a longitudinal temperature gradient $\nabla T\parallel \hat{\bf y}$. The wire acts as a local thermal reservoir at $T_0=(T_L+T_R)/2$. }
\label{model}
\end{figure}

In this Letter, we predict a magnon layer-Hall transport rooted in dipolar scattering, a mechanism distinct from the conventional band-geometry paradigm in electronic systems. As illustrated in Fig.~\ref{model}, we develop a scattering theory for a hybrid chiral architecture comprising a magnetic nanowire atop a high-quality magnetic slab, coupled exclusively via the long-range dipolar coupling. 
When the wire magnon exerts a coupling constant $g_{\kappa}(k_z)$ to the film magnon of wave vector $\kappa\hat{\bf y}+k_z\hat{\bf z}$, we derive the generic condition for the emergence of layer-Hall transport: $|g_{\kappa}(k_z)|\ne |g_{-\kappa}(k_z)|$ and $|g_{\kappa}(k_z)|\ne |g_{\kappa}(-k_z)|$. This potential deflects magnons perpendicular to the temperature gradient, generating opposing transverse thermal Hall currents in the wire and slab. Our analysis reveals a substantial transverse thermal Hall current with a sizable Hall angle of $\sim 6^{\circ}$, notably larger than typical values reported for conventional magnon Hall effects~\cite{Onose2010}. Crucially, both the magnitude and direction of this transverse flow are flexibly tunable via the orientations of the external magnetic field and applied temperature gradient. This robust Hall response in conventional magnetic heterostructures eliminates the need for complex topological engineering and facilitates experimental verification in available experimental setups~\cite{Cosset-Cheneau2024,Baumgaertl2023,Mucchietto2024,Han2021}.

The hybrid system in Fig.~\ref{model} consists of a magnetic film/slab (thickness $s\sim 10$~nm) sandwiched between two thermal reservoirs at temperatures $T_L$ and $T_R$, and a magnetic wire (thickness $d$, width $w\sim 10$~nm) centered atop it. The length of the wire $L_z$ and the length of the slab $L_y$ are in the micrometer range, ensuring that magnon transport occurs in the ballistic regime. An external magnetic field ${\bf H}_0$ at an angle $\theta$ with respect to the wire $z$-axis is applied in the slab $y$-$z$ plane, aligning the saturation magnetization ${\bf M}_s$ of the slab. The saturation magnetization $\tilde{\bf M}_s$ of the wire is independently biased to an angle $\tilde\theta$ with the wire $z$-axis. The magnetic nanowire serves as a local reservoir at temperature $T_0=(T_L+T_R)/2$, which injects magnons into the slab and receives magnons from the slab.

Slab magnons $\hat{a}_{\kappa}(k_z)$ propagate in the $y$-$z$ plane with wave vector ${\kappa}\hat{\bf y}+k_z\hat{\bf z}$, while wire magnons $\hat{a}_l(k_z)$ propagate strictly along the wire $\hat{\bf z}$-direction with wave vector $k_z\hat{\bf z}$.
They are described by the free Hamiltonian $\hat{H}_0=\hbar\sum_{\kappa}\sum_{k_z}\omega_{\kappa}(k_z)\hat {a}^{\dagger}_{\kappa}(k_z)\hat{a}_{\kappa}(k_z)+\hbar\sum_{k_z}\Omega(k_z)(1-i\alpha_G)\hat{a}_l^{\dagger}(k_z)\hat{a}_l(k_z)$,
where $\omega_{\kappa}(k_z)$ is the dispersion of film magnons, $\Omega(k_z)=\sqrt{\omega_1(k_z)\omega_2(k_z)}$ is the dispersion of wire magnons with $\omega_{1,2}(k_z)$ defined in the Supplementary Material (SM)~\cite{supplement}, and $\alpha_G$ parametrizes the damping coefficient of wire magnons. 
The interlayer dipolar interaction mediates magnon coupling via $\hat{H}_{\rm int}=\hbar\sum_{\kappa,k_z}(g_{\kappa}(k_z)\hat{a}_l^{\dagger}(k_z)\hat{a}_{\kappa}(k_z)+g^*_{\kappa}(k_z)\hat{a}^{\dagger}_{\kappa}(k_z)\hat{a}_l(k_z))$, which is a Fano-Anderson Hamiltonian that is widely encountered in condensed matter and atom physics. Here, the coupling strength 
\begin{widetext}
\begin{align}
    g_{\kappa}(k_z)&=-2\mu_0\gamma L_z\sqrt{\tilde M_sM_s}\frac{(1-e^{-kd})(1-e^{-ks})}{k^2\kappa}\sin\left({\frac{\kappa w}{2}}\right)\nonumber\\
    &\times ({\cal M}_{x'}, {\cal M}_{y'})\left(\begin{array}{cc}
        k & -i(\kappa\cos{\tilde\theta}-k_z\sin{\tilde\theta}) \\
        i(k_z\sin{\theta}-\kappa\cos{\theta}) & {(k_z\sin{\theta}-\kappa\cos{\theta})(\kappa\cos{\tilde\theta}-k_z\sin{\tilde\theta})}/{k}
    \end{array}\right)\left(\begin{array}{c}
         \tilde{\cal M}^{*}_{\tilde x}(k_z)  \\
         \tilde{\cal M}^{*}_{\tilde y}(k_z)
    \end{array}\right)
\end{align} 
\end{widetext}
governs the tunneling of magnons between wire and slab. 
Here, ${\cal M}_{x'}=-{1}/(2\sqrt{L_yL_zs})$ and ${\cal M}_{y'}=i{\cal M}_{x'}$ are the normalized amplitudes of film magnons, and $\tilde{\cal M}_{\tilde x}(k_z)=-{[\omega_1(k_z)/\omega_2(k_z)]^{1/4}}/({2\sqrt{L_zws}})$ and $\tilde{\cal M}_{\tilde y}(k_z)=-i{[\omega_2(k_z)/\omega_1(k_z)]^{1/4}}/({2\sqrt{L_zws}})$ are those for wire magnons.
Crucially, this dipolar coupling is intrinsically non-reciprocal, satisfying $|g_{\kappa}(k_z)|\ne |g_{-\kappa}(k_z)|$ and $|g_{\kappa}(k_z)|\ne |g_{\kappa}(-k_z)|$.

The interlayer dipolar interaction hybridizes the magnon modes $\{{\hat a}_{\kappa}(k_z),{\hat a}_l(k_z)\}$ of the film and wire, giving rise to the new modes~\cite{Mahan} (see SM~\cite{supplement})
\begin{align}
    \hat{b}_{\kappa}(k_z,t)&=T_{\kappa,l}(k_z)\hat a_l(k_z)e^{-i\Omega(k_z)t}\nonumber\\
    &+\sum_{\kappa'}T_{\kappa,\kappa'}(k_z){\hat a}_{\kappa'}(k_z)e^{-i\omega_{\kappa'}(k_z)t}, \nonumber\\
    \hat{b}_l(k_z,t)&=\sum_{\kappa}T_{l,\kappa}(k_z){\hat a}_{\kappa}(k_z)e^{-i\omega_{\kappa}(k_z)t},
    \label{output_input}
\end{align}
where the hybridization coefficients are
\begin{align}
&T_{\kappa,l}(k_z)=-\frac{g_{\kappa}^*(k_z)}{\Omega(k_z)(1-i\alpha_G)+\Sigma_{\rm ret}(\omega_\kappa(k_z))-\omega_{\kappa}(k_z)+i\varepsilon},\nonumber\\
&T_{l,\kappa}(k_z)=G(\omega_{\kappa}(k_z))g_{\kappa}(k_z),\nonumber
\end{align}
\begin{align}
&T_{\kappa,\kappa'}(k_z)=\delta_{\kappa,\kappa'}+\frac{g_{\kappa}^*(k_z)G(\omega_{\kappa'}(k_z))g_{\kappa'}(k_z)}{\omega_{\kappa'}(k_z)-\omega_{\kappa}(k_z)+i\varepsilon}.  
\label{Tkappal}
\end{align}
Here, $\varepsilon\rightarrow0^+$ and the retarded Green's function for the wire magnon reads
\[G(\omega_{\kappa}(k_z))=\frac{1}{\omega_{\kappa}(k_z)-\Omega_l(k_z)(1-i\alpha_G)-\Sigma_{\rm ret}(\omega_{\kappa}(k_z))}.\] 
The self-energy  $\Sigma_{\rm ret}(\omega)=\Delta(\omega)-({i}/{2})\Gamma(\omega)$ comprises an induced frequency shift $\Delta(\omega_{\kappa}(k_z))=\sum_{\kappa'}{\cal P}\frac{|g_{\kappa'}(k_z)|^2}{\omega_{\kappa}(k_z)-\omega_{\kappa'}(k_z)}$ and an additional damping with rate $ \Gamma(\omega_{\kappa}(k_z))=\frac{L_y}{|v_y({\kappa},k_z)|}(|g_{\kappa}(k_z)|^2+|g_{-\kappa}(k_z)|^2)$, which characterizes the decay of wire magnons into the continuum of film modes.

The hybridized modes $\{\hat{b}_{\kappa}(k_z,t),\hat{b}_l(k_z,t)\}$ govern the spin and energy transport within the heterostructure. Under the temperature gradient $T_L>T_R$,  magnons injected by the hot reservoir occupy the higher energy and propagate with positive wave vector $+|\kappa|\hat{\bf y}$, while the magnons injected from the cold reservoir occupy the lower energy with negative $-|\kappa|\hat{\bf y}$, forming a thermal current. 
With the detailed derivation given in SM Sec.~III~\cite{supplement}, we obtain the longitudinal and transverse magnon current densities.

The longitudinal ($y$-component) current density in the film is given by
\begin{widetext}
\begin{align}
 J_y&=\frac{1}{L_zL_y}\sum_{k_z}\Bigg\{\sum_{\kappa>0}v_y(\kappa,k_z)\left[f(\omega_{\kappa}(k_z),T_L)-f(\omega_{\kappa}(k_z),T_R)\right]
 \nonumber\\
&-v_y(\kappa({\Omega}),k_z)\frac{|g_{\kappa(\Omega)}(k_z)|^2|g_{-\kappa(\Omega)}(k_z)|^2}{|g_{\kappa(\Omega)}(k_z)|^2+|g_{-\kappa(\Omega)}(k_z)|^2+\alpha_G\Omega(k_z)|v_y({\kappa(\Omega)},k_z)|/L_y}\frac{L_y^2}{|v_y({\kappa(\Omega)},k_z)|^2}\left[f(\Omega(k_z),T_L)-f(\Omega(k_z),T_R)\right]\nonumber\\
 &+v_y(\kappa({\Omega}),k_z)\frac{|g_{-\kappa(\Omega)}(k_z)|^2}{|g_{\kappa(\Omega)}(k_z)|^2+|g_{-\kappa(\Omega)}(k_z)|^2+\alpha_G\Omega(k_z)|v_y({\kappa(\Omega)},k_z)|/L_y}\left[f(\Omega(k_z),T_R)-f(\Omega(k_z),T_0)\right]\nonumber\\
&+v_y(\kappa({\Omega}),k_z)\frac{|g_{\kappa(\Omega)}(k_z)|^2}{|g_{\kappa(\Omega)}(k_z)|^2+|g_{-\kappa(\Omega)}(k_z)|^2+\alpha_G\Omega(k_z)|v_y({\kappa(\Omega)},k_z)|/L_y}\left[f(\Omega(k_z),T_0)-f(\Omega(k_z),T_L)\right]\Bigg\}, 
\end{align}
\end{widetext}
where $v_y(\kappa,k_z)={\partial\omega(\mathbf{k})}/{\partial \kappa}$ represents the group velocity of film magnon along  $\hat{\bf y}$. The wire reservoir at $T_0$ ensures a vanishing longitudinal current when $T_L=T_R=T_0$, thereby obeying the second Law of thermodynamics even when the coupling is nonreciprocal~\cite{Yu2024b}. This implies that, in the linear-response regime with small bias $\Delta T=T_L-T_R$, the longitudinal response is reciprocal with $J_y(\Delta T)=-J_y(-\Delta T)$, while nonreciprocity comes from the dissipation~\cite{Han2021,Cosset-Cheneau2024}.

The transverse ($z$-component) magnon current densities in the film and wire are
\begin{align}
    {J}_z&=-\frac{1}{2L_zL_y}\sum_{k_z}v_z(k_z)\eta(\kappa_{\Omega},k_z)\nonumber\\
    &\times[f(\Omega(k_z),T_L)-f(\Omega(k_z),T_R)], 
    \label{J_zm}\\
    {\tilde J}_z&=\frac{1}{2L_zw}\sum_{k_z}\tilde{v}_{z}(k_z)\eta(\kappa_{\Omega},k_z)\nonumber\\
    &\times[f(\Omega(k_z),T_L)-f(\Omega(k_z),T_R)], \label{tildeJ_zm}
\end{align}
where $\kappa_{\Omega}>0$ satisfies $\omega_{\kappa_{\Omega}}(k_z)=\Omega(k_z)$, and $v_z(k_z)={\partial\omega(\mathbf{k})}/{\partial k_z}$ and $\tilde v_z(k_z)=\partial \Omega(k_z)/\partial k_z$ are the group velocities of magnons in the film and wire, respectively. The ``chirality ratio" 
\[
\eta(\kappa,k_z)\equiv \frac{|g_{\kappa}(k_z)|^2-|g_{-\kappa}(k_z)|^2}{|g_{\kappa}(k_z)|^2+|g_{-\kappa}(k_z)|^2+\alpha_G\Omega(k_z)|v_y({\kappa},k_z)|/L_y}
\]
emerges for $|g_{\kappa}(k_z)|\ne |g_{-\kappa}(k_z)|$. Crucially, a net transverse Hall current requires an additional condition, $|g_{\kappa}(k_z)|\ne |g_{\kappa}(-k_z)|$, to ensure that the $k_z$-summation yields a finite contribution. The net transverse current in the wire and film is counter-propagating, satisfying $J_zL_y\approx -\tilde{J}_zw$ when the film and wire are taken to be the same material. We term this phenomenon a magnon LHE, in analogy to the electron LHE~\cite{Gao2021,Xu2024,Chen2024,Dai2022,Peng2023,Zhang2023,Zhang2024,Feng2023,Liu2024,Tao2024}, as it features layer-resolved counter-propagating transverse flows.

In the chirality ratio $\eta(\kappa,k_z)$,  only those terms in $|g_{\kappa}(k_z)|^2$ that are odd in both $\kappa$ and $k_z$ contribute to the Hall current density. We separate these terms according to 
$|g_{\kappa}(k_z)|^2= \sum_{n=0}^{\infty}\sum_{m=0}^{\infty} c_{mn}\,\kappa^{2n+1} k_z^{2m+1}+\text{(other terms)}$,
where $\{n,m\} = \{0,1,2,\dots\}$ are integers. By further calculation, 
$\sum_{n=0}^{\infty}\sum_{m=0}^{\infty}c_{mn}\kappa^{2n+1}k_z^{2m+1}=W(\kappa,k_z)F(\theta,\tilde\theta,\kappa,k_z)$,
in which $W(\kappa,k_z)$ is an even function of both $\kappa$ and $k_z$ and, with $\tilde\theta\approx\theta$ for the stable states (see Fig.~\ref{different_situations_current} below),
\begin{align}
&F(\theta,\kappa,k_z)\approx-\Big[k^2|{\cal M}_{x'}|^2|\tilde {\cal M}_{\tilde y}(k_z)|^2+k^2|{\cal M}_{y'}|^2|\tilde {\cal M}_{\tilde x}(k_z)|^2\nonumber\\&+4k^2{\cal M}_{x'}{\cal M}_{y'}\tilde{\cal M}_{\tilde x}(k_z)\tilde{\cal M}_{\tilde y}(k_z)+2|\tilde{\cal M}_{\tilde y}(k_z)|^2|{\cal M}_{y'}|^2\nonumber\\
&\times(\kappa^2\cos^2{\theta}+k_z^2\sin^2{\theta})\Big]\kappa k_z\sin{2\theta}\propto\sin{2\theta}
\end{align}
is an odd function of both $\kappa$ and $k_z$. Thereby, the transverse Hall current densities in the wire and film $\{\tilde{J}_z,J_z\}\propto\sin{2\theta}$.  The angular dependence ($\propto\sin{2\theta}$) reflects mirror-symmetry breaking rather than time reversal. This distinguishes the present scattering-driven LHE from conventional Hall effects where the response is odd under time reversal.

We note that magnons in a single-layer ferromagnetic film exhibit no intrinsic Berry curvature when the saturation magnetization lies in-plane~\cite{Matsumoto2011}. This absence persists in our hybrid architecture: both the bilayer subsystem and the coupled wire-slab heterostructure possess vanishing Berry curvature due to a parity--effective-time ($\mathcal{PT}_x^s$) symmetry preserved in the magnon Hamiltonian~\cite{Cheng2016,Mook2019,Zhang2019,Zhou2026} (see SM~\cite{supplement}). Here, ${\mathcal T}^s_{x}$ denotes effective time reversal, namely, time reversal followed by a $\pi$ rotation of magnetization about the normal $x$-axis. Consequently, momentum-space Berry curvature is precluded as the origin of the Hall effect in this system.

We now quantify the magnitude of the layer Hall transport. The hybrid structure consists of a YIG film ($L_y=L_z=10~\mu$m, $s = 10$~nm) topped by a YIG nanowire ($L_z=10~\mu$m, $w=40$~nm, $d = 10$~nm), with $\mu_0M_s=\mu_0\tilde{M_s}=0.21$~T. The temperatures of the reservoirs are set to $T_L = 11$~K and $T_R = 10$~K, with $T_0 = (T_L + T_R)/2=10.5$~K. Since the magnon transport is highly sensitive to the magnetization alignment, we investigate two distinct magnetic configurations realized with different magnitudes of external fields $H_0$, corresponding to stable and metastable states, respectively.

When biased by a relatively strong applied field  $\mu_0 H_0 = 0.05$~T, only the stable state exists: the film magnetization is parallel to the applied field $\theta$ while the wire magnetization is approximately aligned (see the orientation of $\tilde {\bf M}_s$ in the SM~\cite{supplement}). As shown in Fig.~\ref{different_situations_current}(a), the chirality ratio generally exists: it depends on the field direction $\theta$ and is approximately an odd function of $k_z$. 
This indicates that the transverse magnon current densities are generally enabled by the longitudinal temperature gradient, in accordance with Eqs.~\eqref{J_zm} and \eqref{tildeJ_zm}.  
As illustrated in Fig.~\ref{different_situations_current}(b)-(c), the transverse Hall currents in both the wire and film are tunable via the orientation of the applied magnetic field.  $\tilde J_z$ flows along $+\hat{\bf z}$ for $\theta \in (0,90^\circ) \cup (180^\circ, 270^\circ)$, and along $-\hat{\bf z}$ for $\theta \in (90^\circ, 180^\circ) \cup (270^\circ, 360^\circ)$. 
Reflecting the field with respect to the $z$‑ or $y$‑axis preserves the magnitude but flips the direction. The film transverse current $J_z$ flows in the opposite direction to $\tilde J_z$, with its magnitude reduced by a factor of $w/L_z$, i.e., $J_z = -\tilde J_z \, w/L_z$. The magnitude of $|\tilde J_z|$ and $|J_z|$ reach their maximum when the applied field $\theta\approx \{60^\circ,120^\circ,240^\circ,300^\circ\}$.

\begin{figure}[htp!]
\centering
\includegraphics[width=0.52\linewidth]{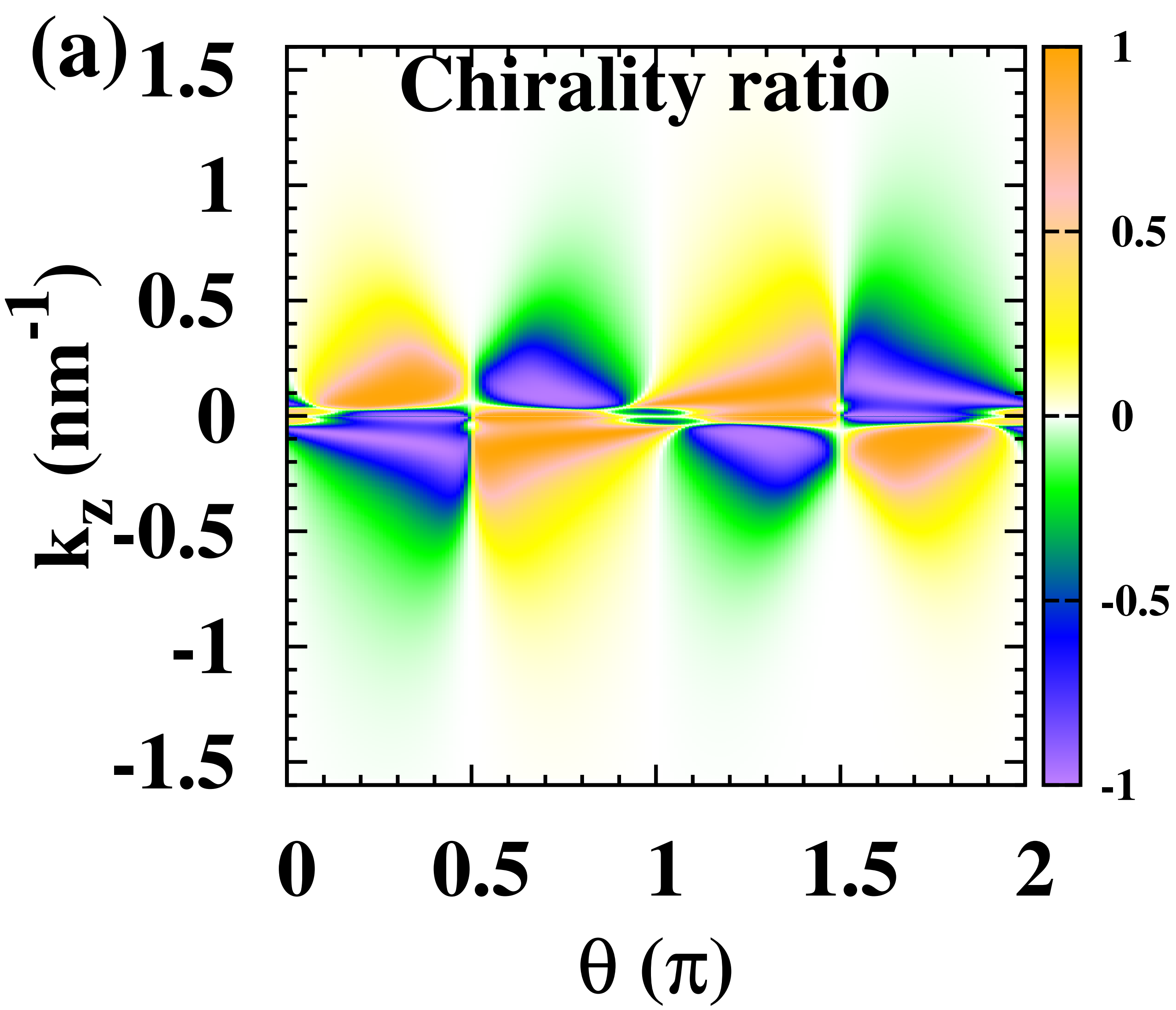}
\hspace{-0.2cm}
\includegraphics[width=0.48\linewidth,trim=0cm -1.9cm 0cm 0cm]{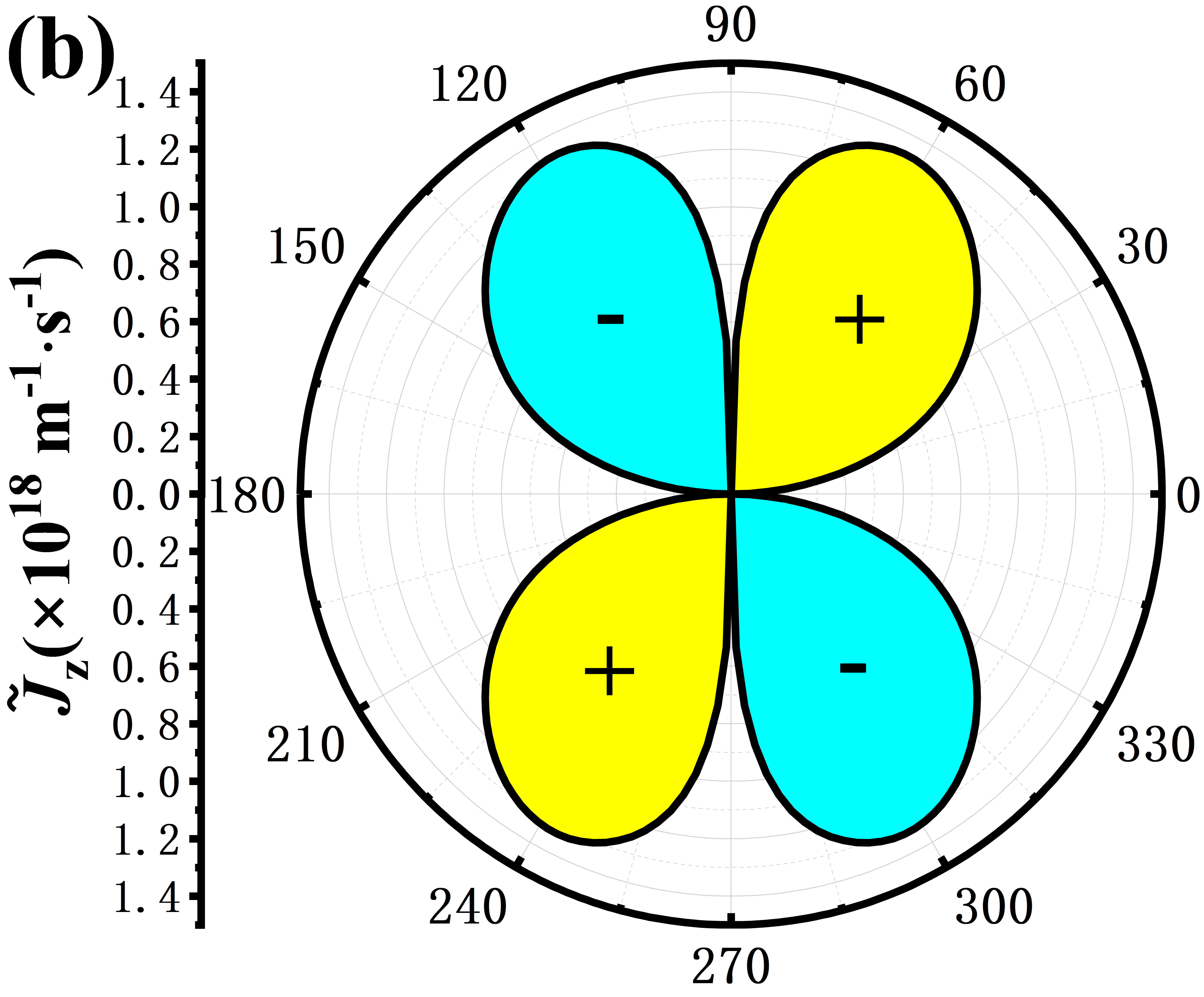}
\hspace{-0.2cm}
\includegraphics[width=0.47\linewidth]{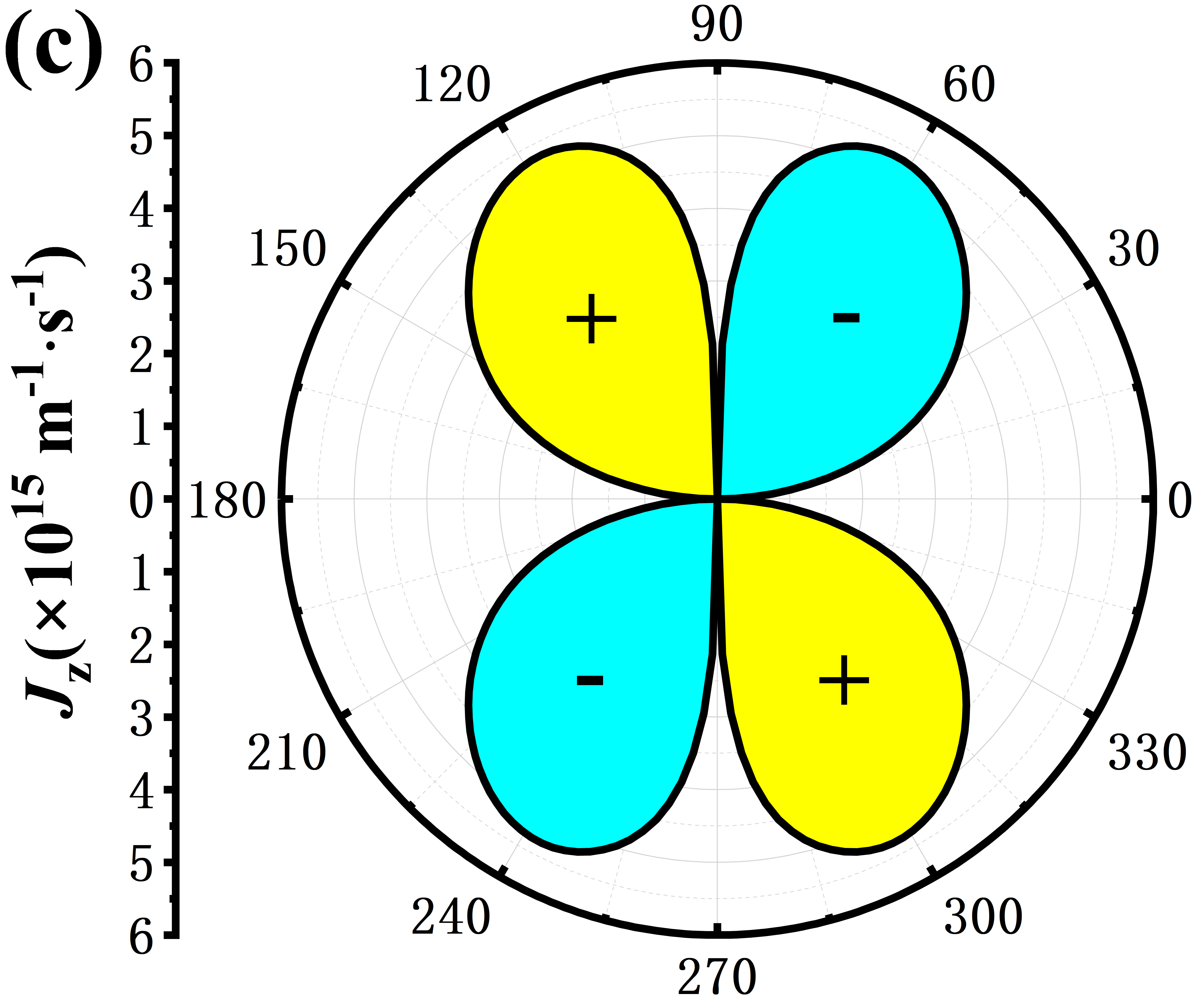}
\hspace{-0.2cm}
\includegraphics[width=0.5\linewidth]{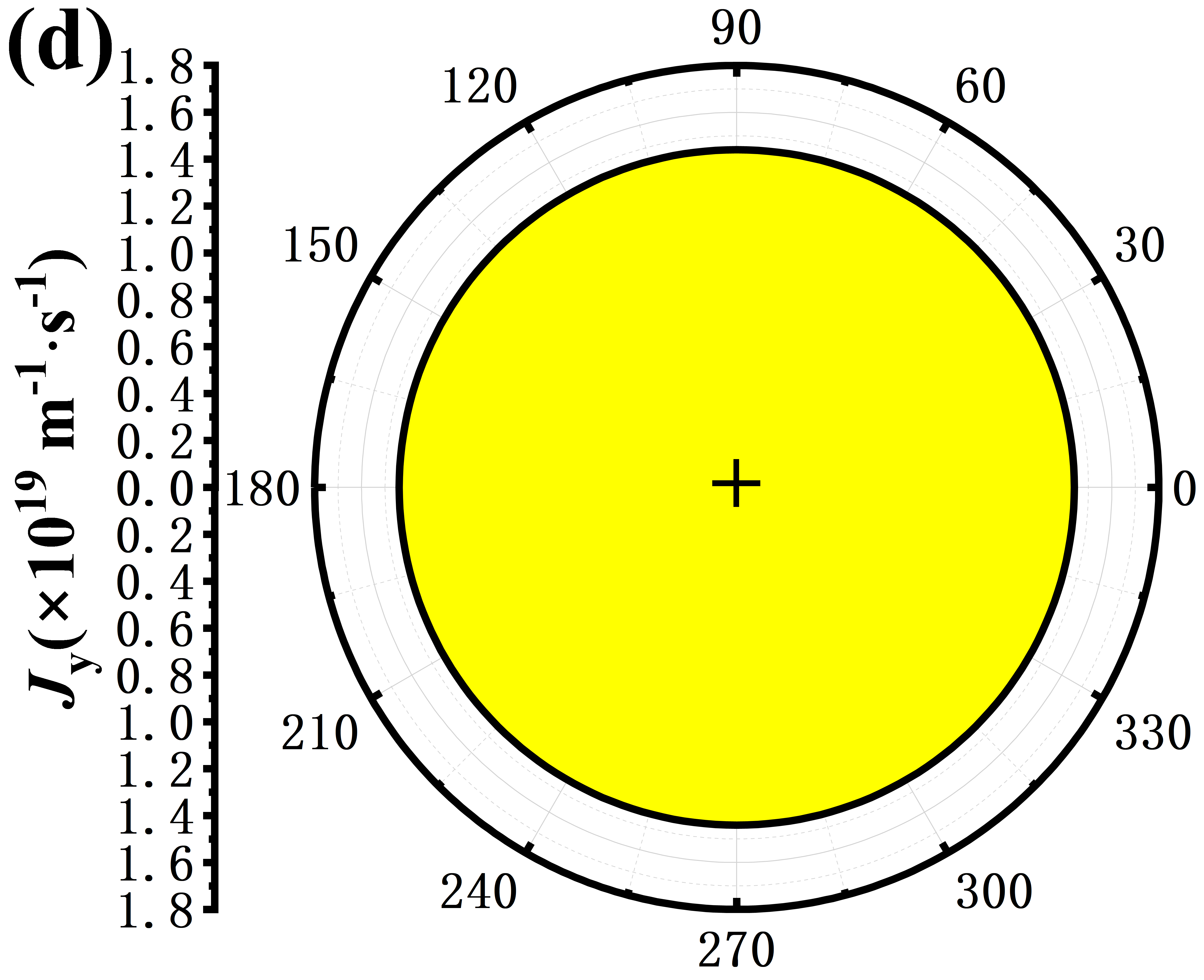}
\hspace{-0.2cm}
\caption{Transverse vs. longitudinal magnon current in the film and wire when biased by an applied field $\mu_0H_0=0.05$~T in different directions, where the system resides in a stable configuration $\tilde{\theta}\approx \theta$.  (a): dependence of chirality ratio $\eta(\kappa_\Omega,k_z)$ on the wave vector $k_z$ and field direction $\theta$. (b) and (c): transverse currents in the wire ($\tilde{J}_z$) and film (${J}_z$) under different magnetization configurations.  (d): longitudinal current ($J_y$) in the film versus the external field direction. }
\label{different_situations_current}
\end{figure}

The longitudinal current $J_y$ in the film is only weakly affected by the dipolar coupling with wire magnons, as shown in Fig.~\ref{different_situations_current}(d). Remarkably, the efficiency $|\tilde J_z / J_y|$ can reach nearly $10\%$, i.e., the Hall angle can reach about $6^\circ$, demonstrating that the effect is sufficiently large for experimental detection~\cite{Onose2010}.

\begin{figure}[htp!]
\centering
\includegraphics[width=0.5\linewidth]{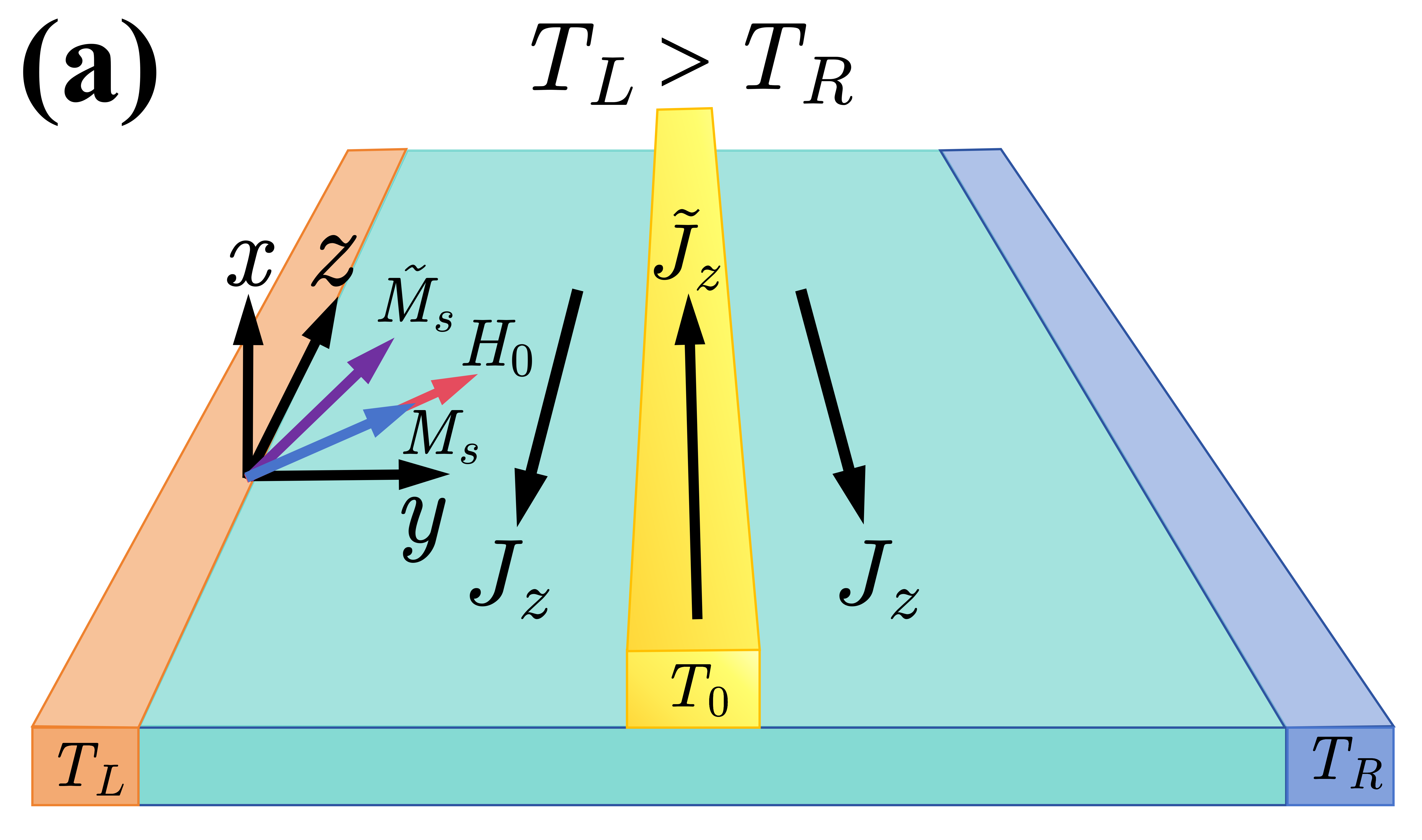}
\hspace{-0.2cm}
\includegraphics[width=0.5\linewidth]{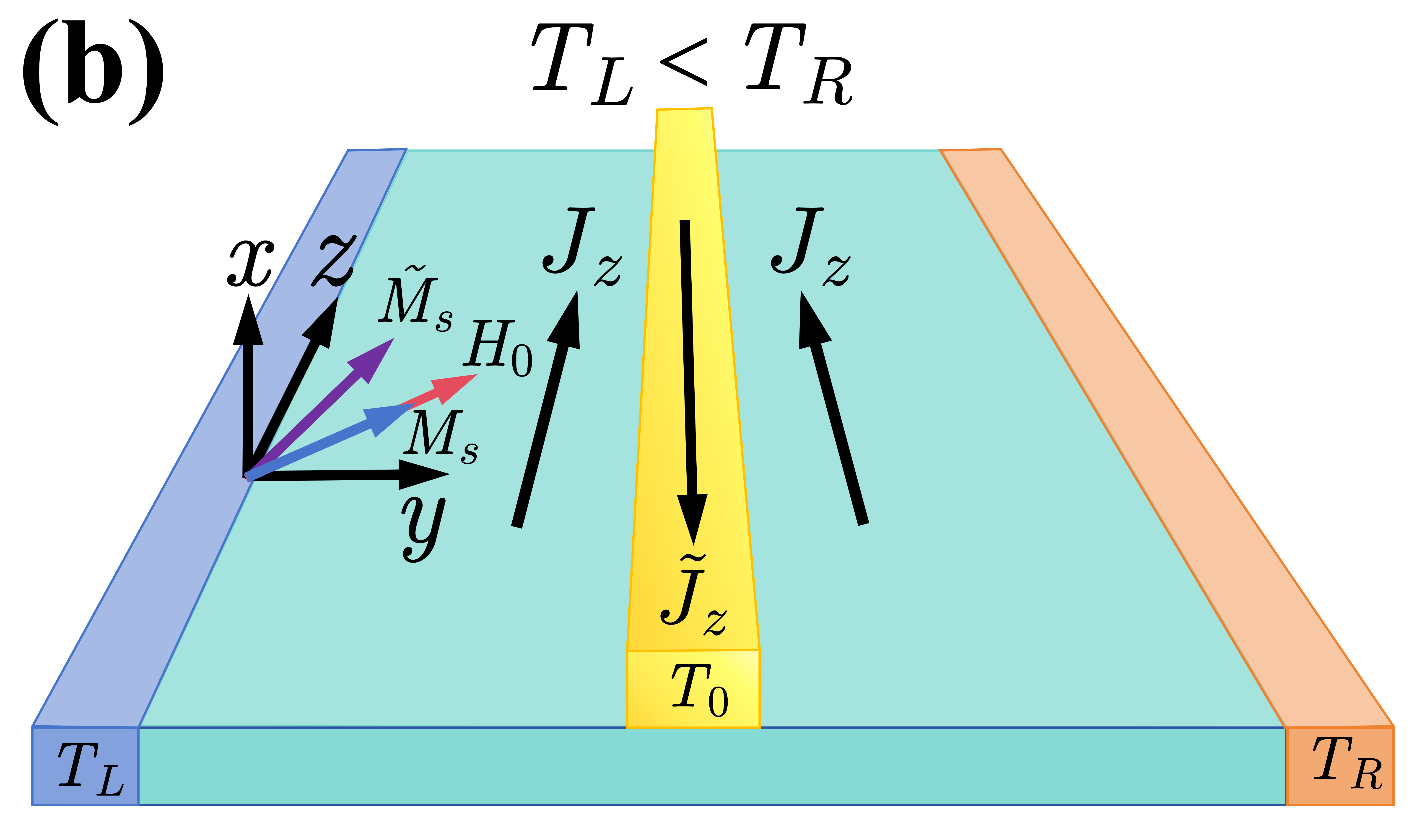}
\hspace{-0.2cm}
\includegraphics[width=0.5\linewidth]{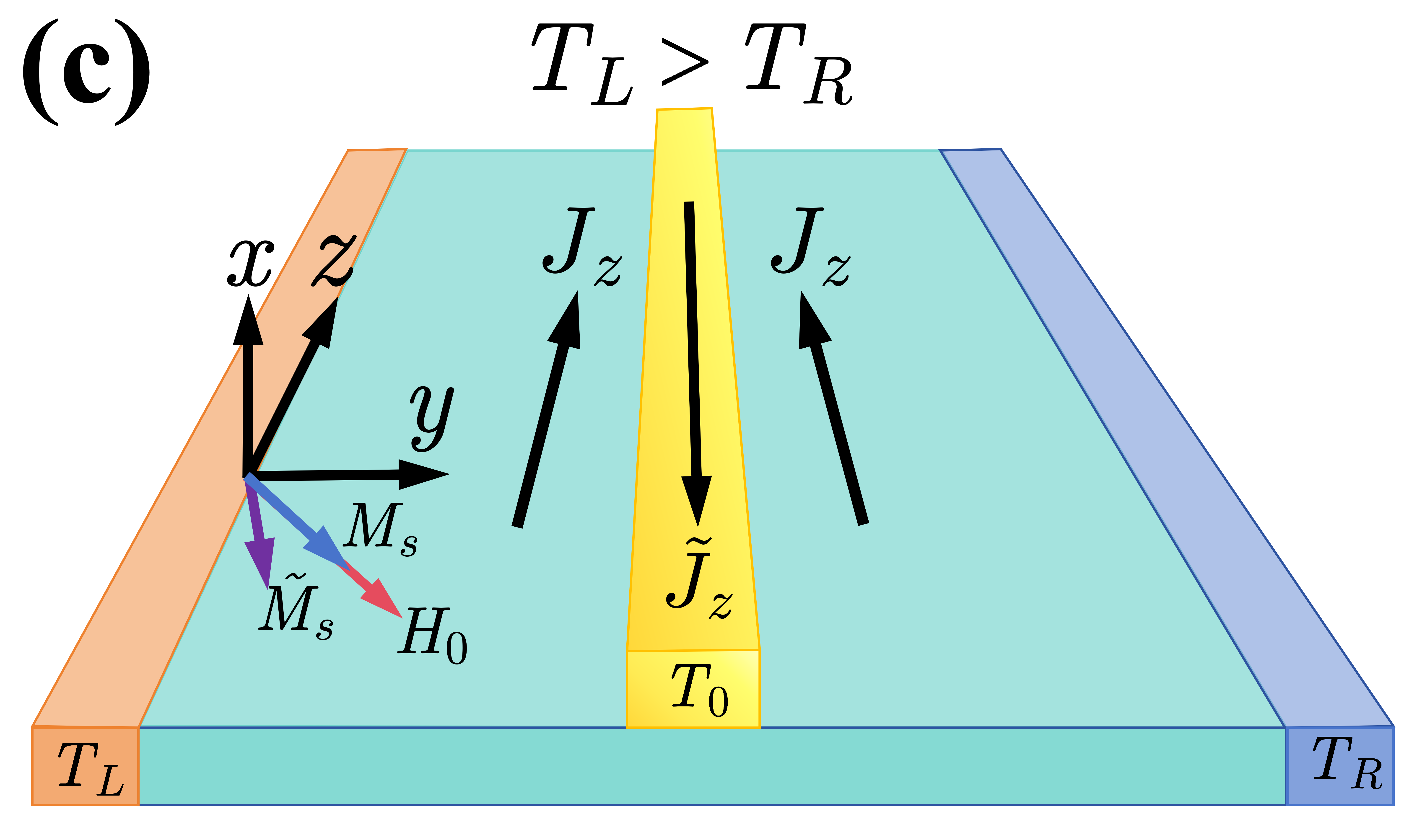}
\hspace{-0.2cm}
\includegraphics[width=0.5\linewidth]{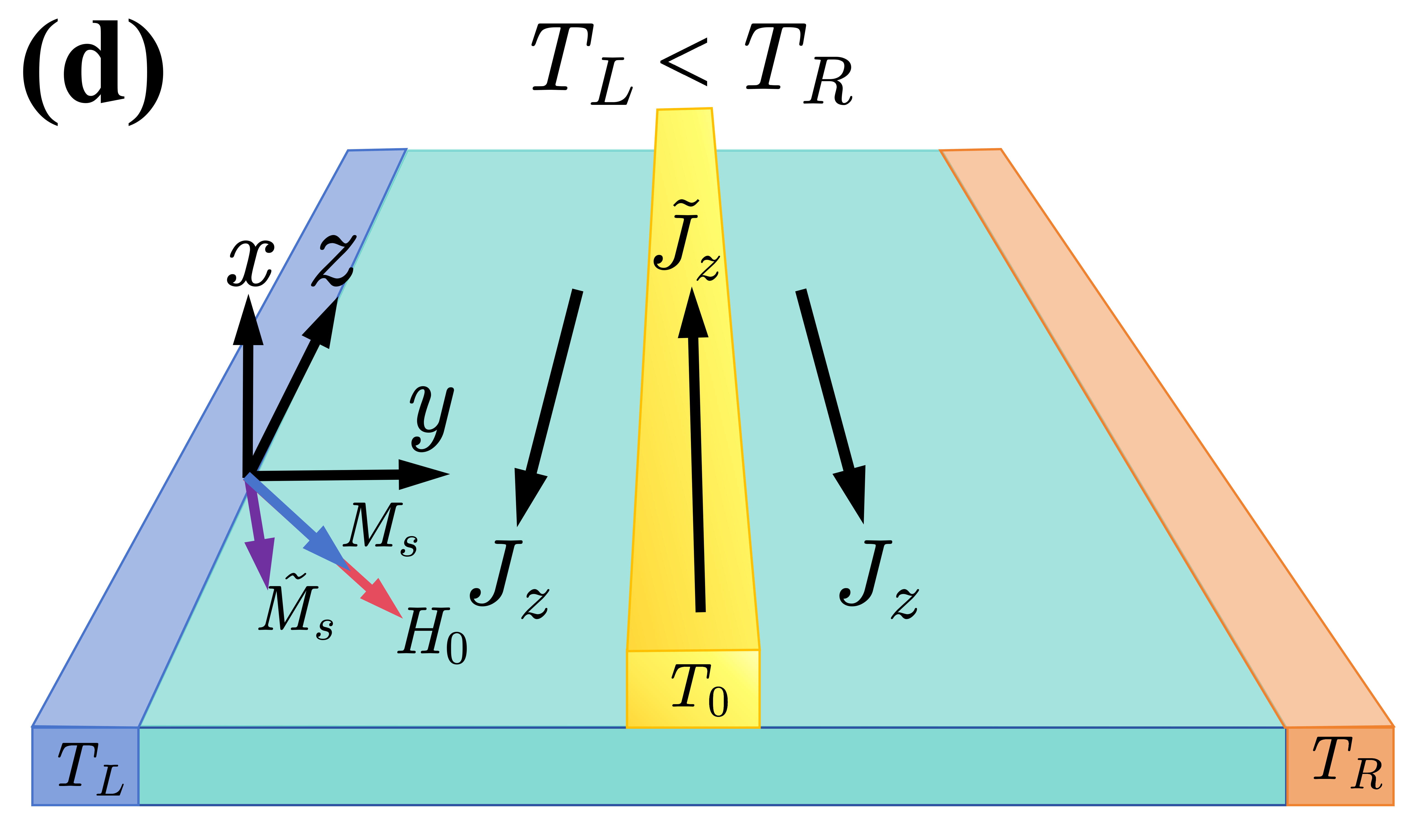}
\hspace{-0.2cm}
\caption{Manipulation of layer Hall current by the temperature gradient direction and the external field ${\bf H}_0$ that changes the directions $\theta$ and $\tilde{\theta}$ of ${\bf M}_s$ and $\tilde{\bf M}_s$. }
\label{different_situations_current2}
\end{figure}

The flow direction is flexibly tunable by the temperature gradient and external-field directions. The wire transverse current approximately satisfies $\tilde J_z \propto -(\mathbf{e}_y\cdot\nabla T)\sin{2\theta}$, while the transverse current in the film flows in the opposite direction. As compared in Fig.~\ref{different_situations_current2}(a) and (b) [(c) and (d)], reversing the thermal gradient reverses the direction of the Hall current. Further, by reflecting the applied magnetic field $H_0$ about the $z$‑axis or the $y$‑axis, the direction of the transverse Hall currents in both the wire and the film is flipped, as comparison in Fig.~\ref{different_situations_current2}(a) and (c) [(b) and (d)].

\begin{figure}[htp!]
\centering
\includegraphics[width=0.52\linewidth]{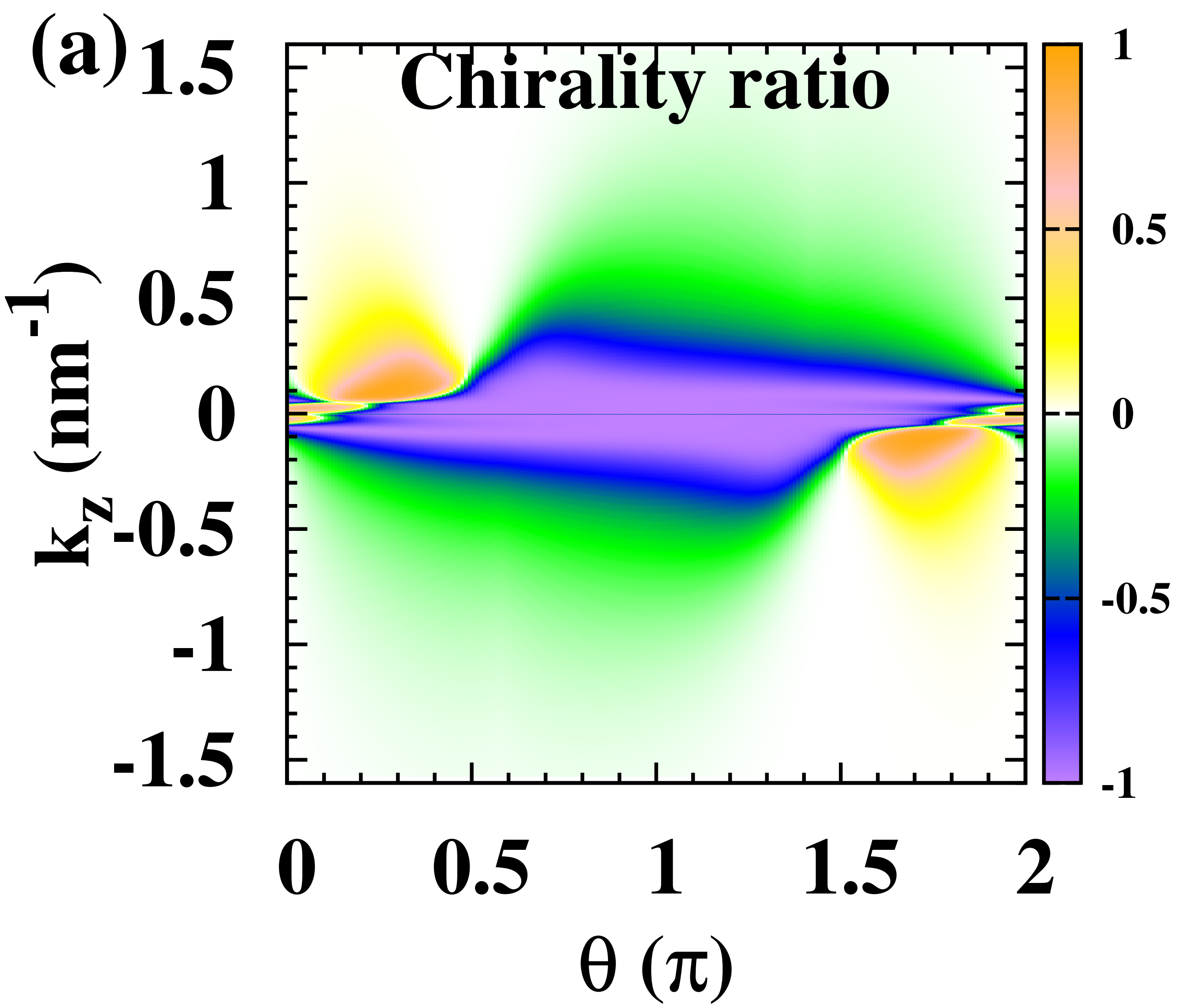}
\hspace{-0.2cm}
\includegraphics[width=0.47\linewidth,trim=0cm -2.4cm 0cm 0cm]{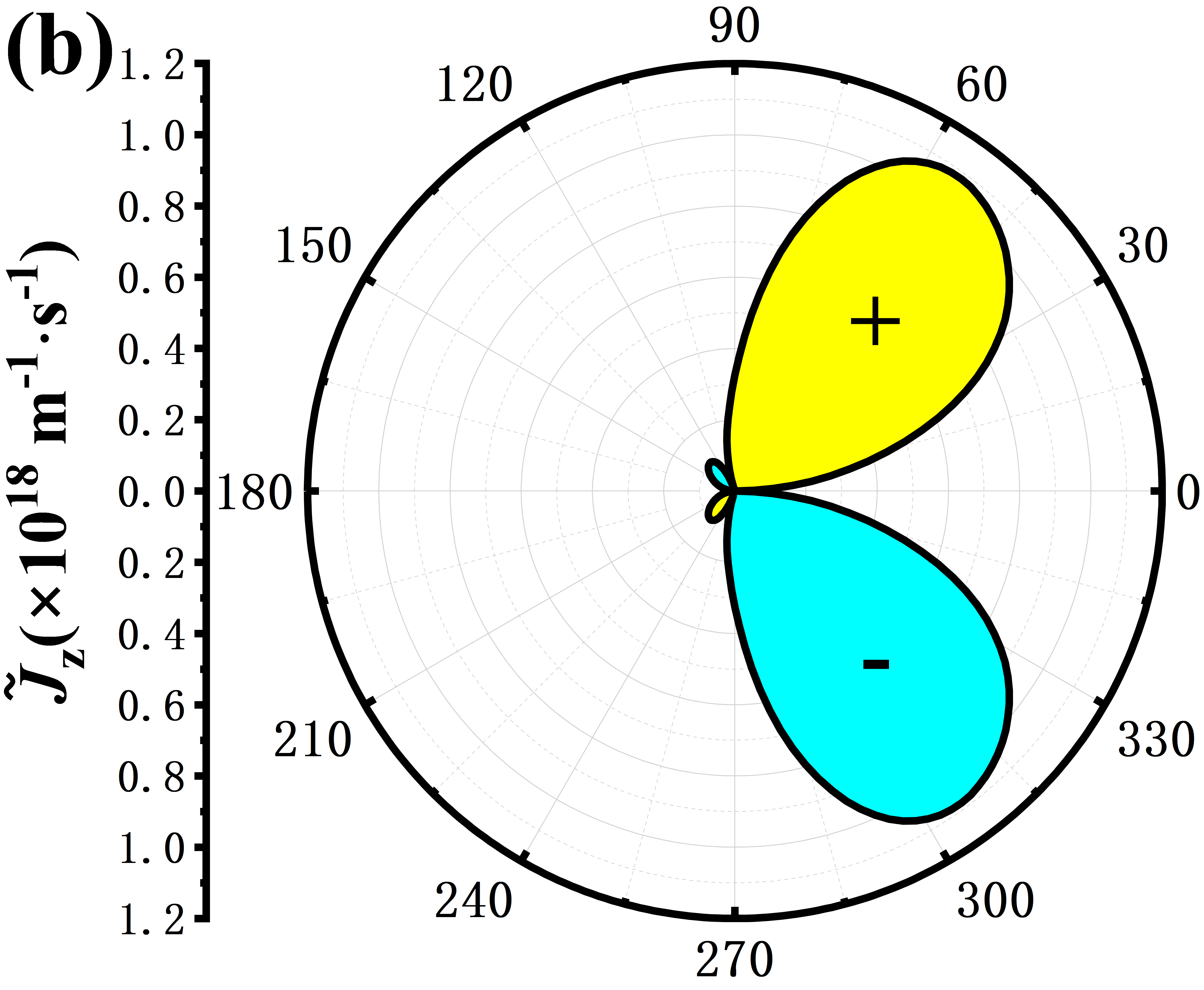}
\hspace{-0.2cm}
\includegraphics[width=0.465\linewidth]{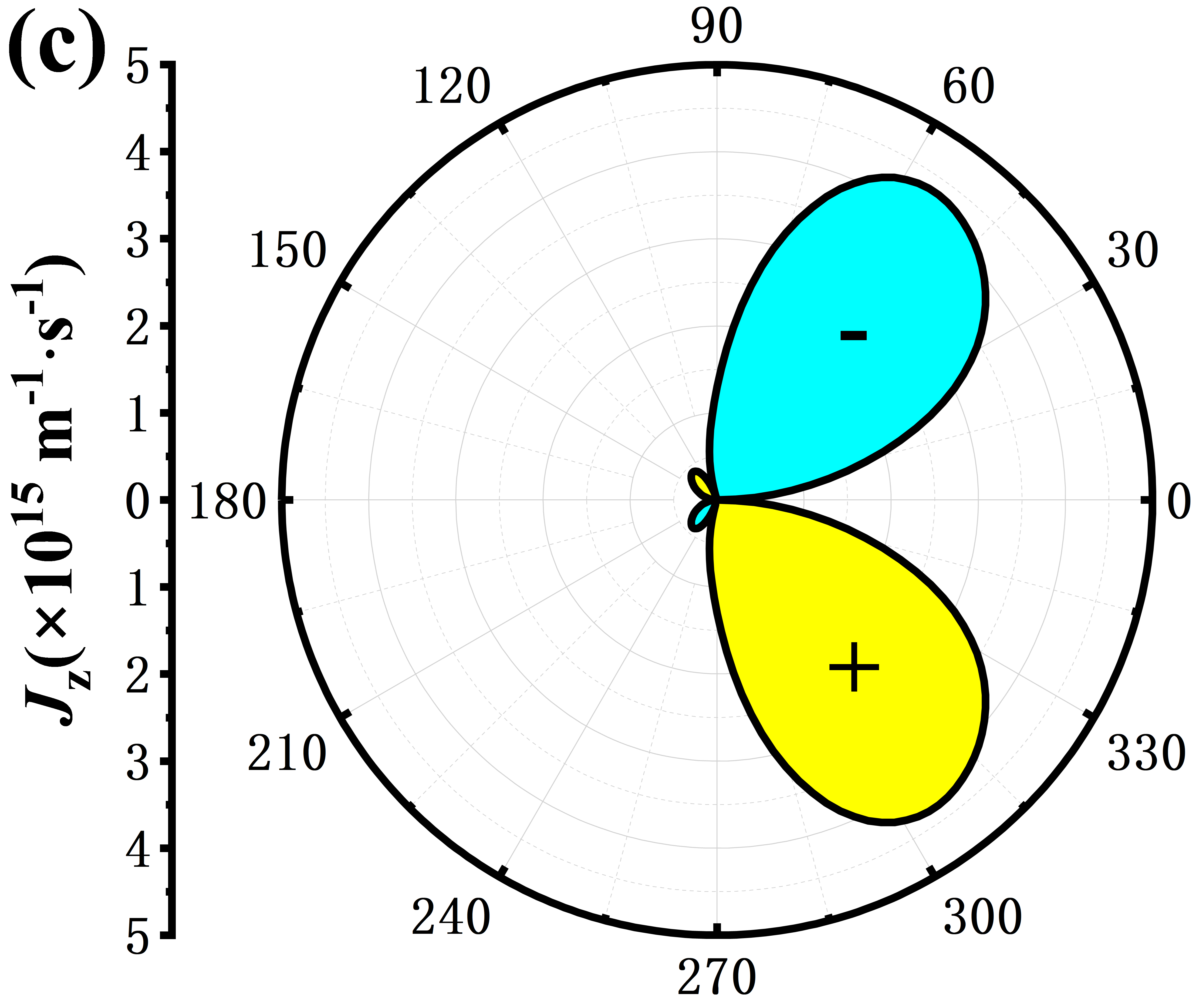}
\hspace{-0.2cm}
\includegraphics[width=0.49\linewidth]{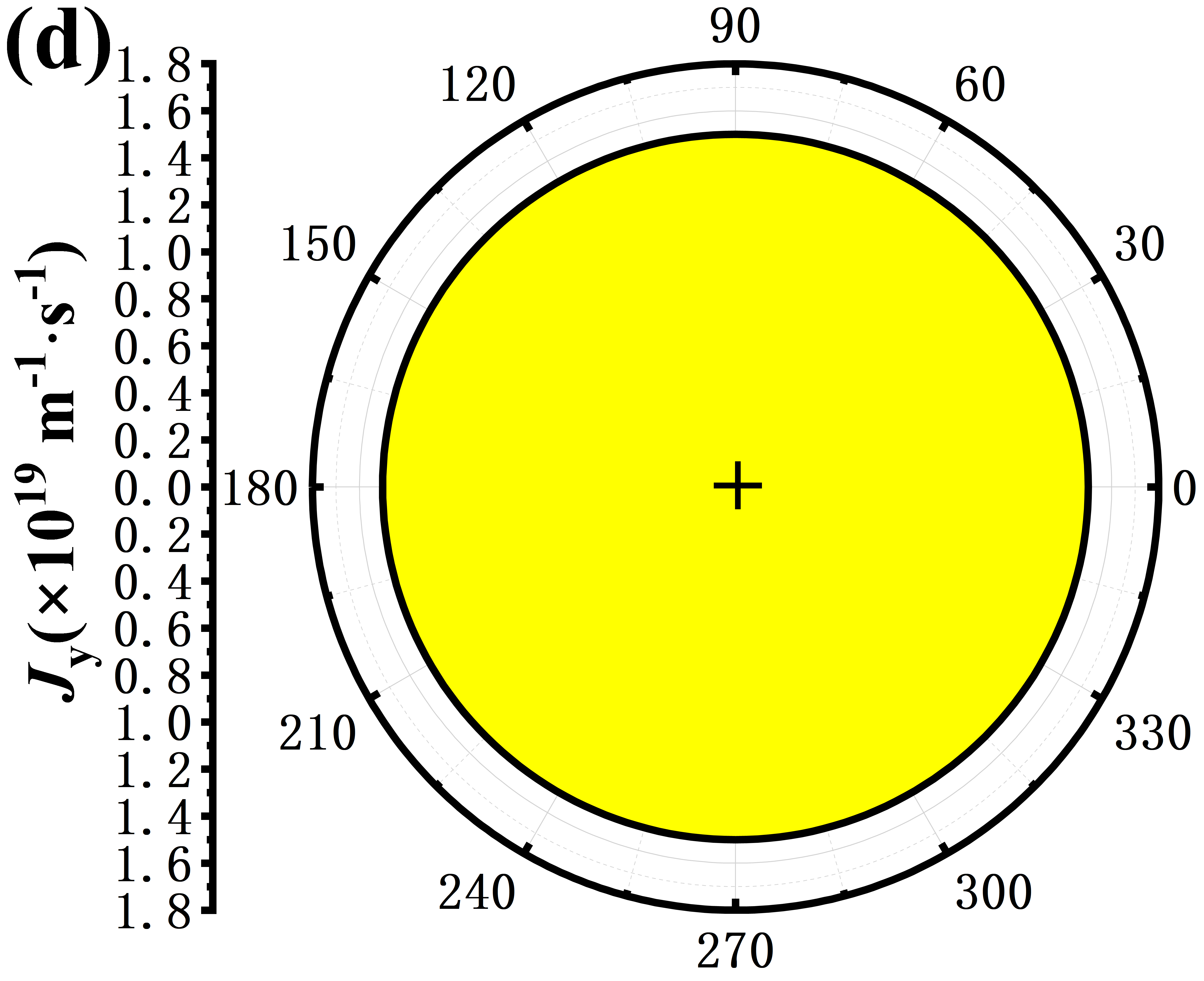}
\hspace{-0cm}
\caption{Magnon current density in the wire and film when biased by a weak applied field $\mu_0H_0=0.01$~T, where metastable states occur when $\theta \in (90^\circ, 270^\circ)$. (a) plots the dependence of the chirality ratio on the wave vector ($k_z$) and the field direction ($\theta$). (b) and (c): switch on and off of the transverse Hall current in the wire $\tilde J_z$ and film $J_z$ by the direction of the external magnetic field. (d): longitudinal magnon current $J_y$ in the film, weakly affected by the external field. }
\label{different_situations_current1}
\end{figure}

The transport under a relatively weak applied field $\mu_0 H_0 = 0.01$~T becomes completely different. The film magnetization is aligned to the applied field, but the wire magnetization is almost fixed to $+\hat{\bf z}$ when $\tilde{\bf M}_s$ is initially along the positive $z$-axis. The saturation magnetization of the wire is metastable when $\theta \in (90^\circ, 270^\circ)$ and is stable otherwise (see SM~\cite{supplement}). 
As illustrated in Fig.~\ref{different_situations_current1}(a),  the chirality ratio is approximately odd in $k_z$ when the field direction $\theta\in(0^\circ, 90^\circ)\cup(270^\circ, 360^\circ)$, signaling the presence of a pronounced Hall current as in Fig.~\ref{different_situations_current1}(b) and (c). By contrast, when $\theta\in(90^\circ, 270^\circ)$, the chirality ratio is nearly even in $k_z$, indicating the switch-off of the Hall current as in Fig.~\ref{different_situations_current1}(b)-(c). 
According to Fig.~\ref{different_situations_current1}(d), the longitudinal magnon current in the film remains weakly affected by the scattering of the wire magnon. The efficiency $|\tilde J_z/J_y|$ reaches nearly $7\%$, confirming that the effect remains well within experimental detectability~\cite{Onose2010}.

The magnon layer-Hall transport induces an accumulation of magnons at opposite transverse edges of the nanowire, a signature that can be probed via infrared thermal imaging~\cite{An2013,Wid2016,Shigematsu2018,PWang2018}. While the longitudinal magnon current in the setup Fig.~\ref{model} has been routinely measured~\cite{Cosset-Cheneau2024,Baumgaertl2023,Mucchietto2024,Han2021}, our theoretical framework suggests that the transverse Hall current is also within experimental reach. Furthermore, the scattering-induced mechanism presented here is not confined to magnons; it extends universally to other quasiparticles, such as ferrons~\cite{Bauer2022,Zhou2023,Choe2026} and polar phonons~\cite{Venkataram2018,Deng2021}, provided that their interlayer dipolar coupling satisfies the requisite symmetry conditions.

In conclusion, we have established a scattering theory for the magnon LHE rooted in a chiral scattering mechanism, distinct from conventional band-geometry paradigms. Our analysis of the wire-film heterostructure reveals a substantial Hall current under generic conditions, with the wire Hall current reaching up to $10\%$ of the film's longitudinal current. Both the magnitude and sign of this transverse response are readily reconfigurable via the orientation of the applied magnetic field and the thermal gradient. This scattering-driven LHE provides a reconfigurable platform for magnon-based thermal logic and sensing without topological engineering.

\begin{acknowledgments}
This work is financially supported by the National Natural Science Foundation of China under Grants No.~12374109 and 12474047, and the National Key Research and Development Program of China under Grant No.~2023YFA1406600. Z.J.S. acknowledges financial support by SUSTech Presidential Postdoctoral Fellowship.
\end{acknowledgments}

\end{document}